\documentclass[aps,preprint,showpacs,amsmath,amssymb]{revtex4}
\usepackage{epsfig}
\usepackage{wasysym}
\usepackage{amssymb}

\begin{document}

\newcommand{\bvec}[1]{\mbox{\boldmath ${#1}$}}
\title{Evidence for the $J^p=1/2^+$ narrow state at
       1650 MeV in the photoproduction of $K\Lambda$}
\author{T. Mart}
\affiliation{Departemen Fisika, FMIPA, Universitas Indonesia, Depok 16424, 
  Indonesia}
\date{\today}
\begin{abstract}
 We have investigated the existence of the $J^p=1/2^+$ 
 narrow resonance predicted by the 
 chiral soliton model by utilizing the 
 kaon photoproduction process $\gamma +p\to K^++\Lambda$.
 For this purpose we have constructed two phenomenological models 
 based on our previous effective Lagrangian model, 
 which are able to describe kaon photoproduction from 
 threshold up to $W=1730$ MeV. By varying the mass (width)
 of an inserted $P_{11}$ resonance from 1620 to 1730 MeV 
 (0.1 to 1 MeV and 1 to 10 MeV) a number of fits has 
 been performed in order to search for the resonance mass.
 Our result indicates that the most promising candidate mass (width)
 of this resonance is 1650 MeV (5 MeV). Although our calculation
 does not exclude the possibility of narrow resonances with masses
 of 1680, 1700 and 1720 MeV, the mass of 1650 MeV  
 is obtained for all phenomenological 
 models used in this investigation. Variations of the 
 resonance width and $K\Lambda$ branching ratio are 
 found to have a mild effect on the $\chi^2$.
 The possibility that the obtained result originates from other resonance
 states is also discussed.
\end{abstract}

\maketitle

\section{Introduction}
\label{sec:intro}
\begin{figure}[t]
  \begin{center}
    \leavevmode
    \epsfig{figure=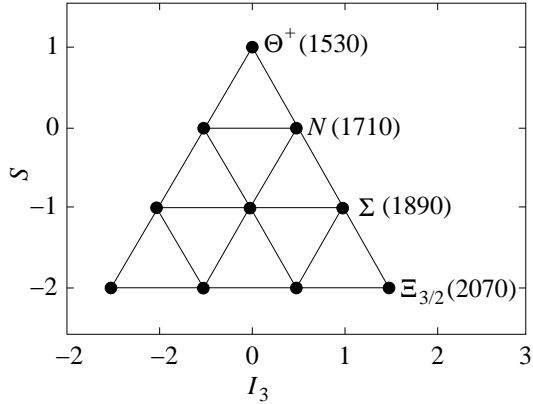,width=70mm}
    \caption{Antidecuplet baryons predicted by the
      chiral soliton model \cite{diakonov}.
      }
   \label{fig:antidecuplet} 
  \end{center}
\end{figure}

The ten members of the antidecuplet baryons predicted by 
the chiral quark soliton model ($\chi$QSM) have drawn
considerable attention for more than a decade. 
According to their strangeness and isospin 
these baryons can be organized as in 
Fig.~\ref{fig:antidecuplet}. Among them three 
are exotic in the sense that their quantum numbers can be only
built from 5 quarks, whereas the simplest states 
are clearly the two nonstrange nucleon resonances with 
$J^p=1/2^+$ indicated by $N(1710)$ in the figure. 
It is interesting to learn that the mass of 1710 MeV
was originally assigned by
Diakonov, Petrov, and Polyakov \cite{diakonov} 
to these nonstrange antidecuplet members 
since at the time the  Particle Data Group (PDG) \cite{barnett} 
reported the resonance partial decay widths similar
to those predicted by the $\chi$QSM, i.e., strong
decay to the $\eta N$ channel, whereas decays to 
the $\pi N$ and $K\Lambda$ channels are relatively small,
but comparable.
Moreover, the total width of the $P_{11}(1710)$ reported 
by the PDG was uncertain \cite{barnett}. 

Immediately after experimental observations of the exotic 
baryons $\Xi_{3/2}$ \cite{NA49} and $\Theta^+$ \cite{nakano} 
had been reported, Walliser and Kopeliovich \cite{Walliser:2003dy} 
found that the mass splitting within the baryon antidecuplet
in Ref.~\cite{diakonov} is overestimated by more than a factor
of 1.5. By taking into account the SU(3) configuration mixing
Walliser and Kopeliovich obtained the mass of 
the $P_{11}$ should be either 1650 MeV or 1660 MeV, depending on
whether a certain symmetry breaker (called $\Delta$ in 
Ref.~\cite{Walliser:2003dy}) is considered or not, respectively.
We note that the agreement with experimental data is significantly
improved if the $\Delta$ symmetry breaker is included in the
calculation. In other words, within the topological soliton model
of Walliser and Kopeliovich, 
experimental data prefer 1650 MeV for the mass of the $P_{11}$.

Not long after the finding of Walliser and Kopeliovich, Diakonov and Petrov 
\cite{diakonov2004}
reevaluated the mass of the $N^*(1710)$ in Ref.~\cite{diakonov} 
by using masses of these  
exotic baryons as inputs. It was found that the mass of
the nonstrange member of this antidecuplet should be 1647 MeV
if the possible mixing with lower-lying nucleonlike 
octet was not considered, but if the mixing was included 
the mass might shift upward to 1690 MeV. 
The width of the narrow resonance $P_{11}$ 
was originally estimated to be 41 MeV  \cite{diakonov}. 
However, from an analysis of the $\pi N$ data, it was suggested 
the existence of a new narrow state $N^*(1680)$ with very small
$\pi N$ branching \cite{igor}.

The large $\eta N$ branching ratio predicted by the $\chi$QSM
has sparked considerable interest in reevaluation of the
$\eta$ photoproduction at energies around 1700 MeV. It was
then reported that the cross section for the production off 
a free neutron is experimentally found to have a substantial 
enhancement at $W\approx 1670$ MeV \cite{kuznetsov}. This result
has been confirmed by experiments of other collaborations
\cite{confirmed}. Such enhancement is absent 
or very weak in the 
case of free proton. Clearly, the enhancement could be explained
as the presence of the narrow $P_{11}$ resonance
\cite{Fix:2007st}. However,
different explanations are also possible.
Within an SU(3) coupled channels model \cite{Doring:2009qr}
the phenomenon can be described as the contributions
of the $K\Lambda$ and $K\Sigma$ loops. Due to the 
cancellation with contributions from other channels, 
this cross section enhancement does not
exist in the $\gamma p\to \eta p$ process. On the other
hand, the Giessen group interpreted this enhancement
as an interference effect between the $S_{11}(1650)$ and 
$P_{11}(1710)$ states \cite{giessen}. The situation
became more complicated as Ref.~\cite{kim} found
that this enhancement could be generated by other 
resonance states with different parities and spins. 

In the $\pi N$ sector there is only one notable
study of this resonance \cite{igor}. In this study 
the narrow $P_{11}$ mass is obtained from $\pi N$ data
by using a modified partial wave analysis (PWA), 
since the standard PWA can miss narrow
resonances with $\Gamma<30$ MeV \cite{igor,arndt:2009}.
The changes in the total
$\chi^2$ were scanned in the range of resonance mass between 1620 to
1760 MeV after the inclusion of this resonance in the $P_{11}$
partial wave. A clear effect was observed at 1680 MeV and
a weaker one was detected at 1730 MeV. The same result 
was always obtained although the total width was varied
between 0.1 and 10 MeV and branching ratio was also
varied between 0.1 and 0.4.

To our knowledge there has been no attempt to study this
resonance by utilizing kaon photoproduction, although  
kaon photoproduction could offer a new arena for 
investigating this problem due to the explicit presence
of strangeness in the final state. As stated above, 
the $N^*\to K\Lambda$ and  $N^*\to \pi N$ branching ratios 
are predicted by the $\chi$QSM 
to be comparable \cite{diakonov}. Partial
wave analysis of the $\pi N$
data yields the value of $\Gamma_{\pi N}=0.5$ MeV,
whereas theoretical analysis based on soliton picture
results in  $\Gamma_{K\Lambda}=0.7$ (1.56) MeV for
$m_{N^*}=1680$ (1730) MeV \cite{igor}. 
 In view of
this we decide to follow the procedure developed in 
Ref.~\cite{igor}, i.e., we shall scan the changes in
the total $\chi^2$ after including a $P_{11}$ narrow 
resonance with the variation of the resonance mass, width,
and $K\Lambda$ branching ratio. Such a procedure is 
apparently 
suitable for kaon photoproduction, since the cross 
sections are relatively much smaller than in the
case of $\pi N$ or $\eta n$, whereas the experimental 
error bars are in general relatively larger. As 
we can see in the next section, it is difficult
to observe a clear structure in the cross sections
at the energy of interest.

In spite of the difficult situation in kaon photoproduction,
the accuracy of phenomenological model used in this
study is crucial. Since the energy of interest is
very close to the $K\Lambda$ threshold, an accurate
model that can describe experimental data at low energies 
would be much better for this purpose, rather 
than a global model that fits to 
a wide energy range but tends to overlook the 
appearing small structures near threshold. Therefore, in
this paper we shall start with the model developed in 
our previous analysis \cite{Mart:2010ch}. Because the
model was constructed to explain experimental data 
only up to 50 MeV above the threshold, an extension
of energy coverage is mandatory. In 
Ref.~\cite{igor} the change of $\chi^2$ was investigated 
up to $W=1760$ MeV. Although we could in principle
take 1760 MeV as the upper limit of our extended model, at
$W\approx 1730$ MeV the problem of data discrepancy,
between SAPHIR and CLAS data, starts 
to appear in the $K\Lambda$ photoproduction \cite{Mart:2006dk}. 
Extending the model beyond $W\approx 1730$ MeV results in 
a large $\chi^2$, which is obviously not suitable for the
present purpose. 
On the other hand, as discussed above, Ref.~\cite{igor}
found that the most convincing mass of the narrow resonance
is 1680 MeV. Therefore, we believe that it is
sufficient to extend the model up to $W=1730$ MeV
and study the interval between threshold
and $W\approx 1730$ MeV. This argument is also supported by 
the fact that no hadronic form factor is required to
explain data up to this point, which is more
favorable since it can simplify the
reaction amplitudes and simultaneously reduce the number
of uncertainties in the model.
As shown in Ref.~\cite{Haberzettl:1998eq}, 
the inclusion of this form factor leads to the problem 
of gauge invariance in the amplitude and, therefore, 
needs a proper treatment for restoring the gauge invariance.

This paper is organized as follows. In Section \ref{sec:model}
we shall discuss the extension of our photoproduction model.
Section \ref{sec:search} presents and discusses the result of
our search for the narrow resonance. In Section \ref{sec:S11},
we give a brief discussion on the possibility that the 
obtained resonance is not a $P_{11}$ state. In Section \ref{sec:summary} 
we summarize our work and conclude our findings. 

\section{Extending the photoproduction model}
\label{sec:model}
\subsection{The model}
Our previous model \cite{Mart:2010ch} was constructed from 
the standard $s$-, $u$-, and $t$-channel Born terms along 
with the $K^{*+}(892)$ and $K_1(1270)$ $t$-channel vector 
mesons. To improve the agreement with experimental data,
an $S_{01}(1800)$ hyperon resonance was also added to
the background amplitude.
In the $s$-channel term only the $S_{11}(1650)$ resonance
state was included, since between threshold and the upper 
energy limit ($W=1660$ MeV) only this resonance may exist.
To facilitate the following discussion we
need to present the corresponding resonant electric 
multipole from our previous analysis, i.e.,
\begin{eqnarray}
  \label{eq:em_multipole}
  E_{0+}(W) &=& {\bar E}_{0+} \, c_{K\Lambda }\, \frac{f_{\gamma R}(W)\, 
    \Gamma_{\rm tot}(W) m_R\, f_{K R}(W)}{m_R^2-W^2-im_R\Gamma_{\rm tot}(W)}~ e^{i\phi} ~,
  \label{eq:m_multipole}
\end{eqnarray}
where ${\bar E}_{0+}=-A_{1/2}^{0+}$, 
$W$ the total c.m. energy, $\Gamma_{\rm tot}$ the total width, 
$m_R$ the physical mass, and $\phi$ the phase angle.
The energy dependent partial width $\Gamma_{K\Lambda }$ 
is related to the single kaon branching ratio $\beta_K$
via $\Gamma_{K\Lambda } = \beta_K\Gamma_R (|\bvec{q}_K|/q_R)
(W_R/W)$, with $\Gamma_R$ and $q_R$ the total width and kaon 
c.m. momentum at $W=m_R$, respectively. 
The explanation of other factors in Eq.~(\ref{eq:m_multipole})
can be found in Ref.~\cite{Mart:2010ch}. 
The available experimental data from threshold
up to $W=1660$ MeV were fitted by adjusting the coupling
constants of the $K^{*+}(892)$, $K_1(1270)$, and $S_{01}(1800)$ 
intermediate states, as well as the phase angle $\phi$ of the 
$S_{11}(1650)$ resonance. Note that the older versions
of SAPHIR data \cite{saphir98}
were omitted from our database since the
latest version \cite{Glander:2003jw} has better statistics
and comes from the same experiment as the older ones.
Furthermore, the leading  
coupling constants were fixed to the SU(3) prediction, 
i.e. $g_{K\Lambda N}/\sqrt{4\pi}=-3.80$ and 
$g_{K\Sigma N}/\sqrt{4\pi}=1.20$, whereas except for the 
resonance phase angle, all resonance parameters of 
the $S_{11}(1650)$ were taken from the PDG values
\cite{nakamura}.

Compared to older analyses of kaon photoproduction, the result 
of the fits showed a much better agreement with experimental 
data considered. It was also found that the pseudoscalar coupling yields
a more significant improvement than the pseudovector one, especially
in the case of the total and differential cross sections. 

For the purpose of the present calculation we have to extend this
model in order to take into account higher energy data since the
$\chi$QSM \cite{diakonov,Walliser:2003dy} predicts the nonstrange 
member of the antidecuplet $N^*$ to have a mass between 
1650 and 1690 MeV. 
However, the latest calculation from the GWU group found that
the most promising candidate mass is 1680 MeV, although another
weak signal at 1730 MeV is not excluded. Therefore, it is 
sufficient to extend our previous model 
up to $W=1730$ MeV. In the energy range between reaction threshold 
and 1730 MeV we observe that there exist 6 nucleon resonances
listed in the recent 
Particle Data Book \cite{nakamura}. Their
properties relevant to the present work are summarized in 
Table~\ref{tab:resonance_pdg}.

\begin{table}[tb]
  \centering
  \caption{Properties of the nucleon resonances taken from the 
  Review of Particle Properties \cite{nakamura}.}
  \label{tab:resonance_pdg}
  \begin{ruledtabular}
  \begin{tabular}[c]{cccccccc}
    Resonance&$M_R$&$\Gamma_R$&$\beta_K$&$A_{1/2} (p)$ &$A_{3/2} (p)$ & Overall& Status\\
             &(MeV)&(MeV)     &         &($10^{-3}$GeV$^{-1/2}$)&($10^{-3}$GeV$^{-1/2}$)& status & seen in $K\Lambda$\\
    \hline
    $S_{11}(1650)$&$1655^{+15}_{-10}$ & $165\pm 20$ &$0.029\pm 0.004$& $+53\pm 16$&-&****&***\\
    $D_{15}(1675)$&$1675\pm 5$ & $150^{+15}_{-20}$ &$<0.01$&$+19\pm 8$&$+15\pm 9$ &****&*\\
    $F_{15}(1680)$&$1685\pm 5$ & $130\pm 10$ &-&$-15\pm 6$&$+133\pm 12$ &****&\\
    $D_{13}(1700)$&$1700\pm 50$ & $100\pm 50$ &$<0.03$&$-18\pm 13$&$-2\pm 24$ &***&**\\
    $P_{11}(1710)$&$1710\pm 30$ & $100^{+150}_{-50}$ &$0.15\pm 0.10$& $+9\pm 22$&-&***&**\\
    $P_{13}(1720)$&$1720^{+30}_{-20}$ & $200^{+100}_{-50}$ &$0.044\pm 0.004$ &$+18\pm 30$& $-19\pm 20$&****&**\\
  \end{tabular}
  \end{ruledtabular}
\end{table}

We also note that there is no resonance state listed in 
the Particle Data Book between 1720 MeV and 1900 MeV. Therefore, we 
are convinced that it is sufficient to extend the model up to
$W=1730$ MeV. It is also important to mention here that the problem
of data discrepancy between SAPHIR \cite{Glander:2003jw}
and CLAS \cite{Bradford:2005pt} data starts to appear
at this energy (see Ref.~\cite{Mart:2006dk} for a thorough
discussion on this problem).

In the extended model we maintain the background terms as in
our previous work \cite{Mart:2010ch}, but in the resonance terms
we include all six resonance states listed in 
Table~\ref{tab:resonance_pdg}. Since the number of experimental 
data considered in the present work (704 points) is much larger 
than that in the previous work (139 points) it is important to
relax the coupling constants restriction in order to achieve
an acceptable $\chi^2$ in our fits. Thus, for instance, we allow the 
main coupling constants to vary within the allowed SU(3) 
values, assuming the SU(3) symmetry is broken at the level 
of 20\%, i.e., $-4.4\le g_{K\Lambda N}/\sqrt{4\pi}\le -3.0$ and 
$0.9\le g_{K\Sigma N}/\sqrt{4\pi}\le 1.3$. At this stage
it is important to note that in the energy of interest 
the hadronic form factors are not required for the 
background terms. This fact reduces the level of uncertainty 
and complexity in our model since the problem of gauge 
invariance due to the inclusion of hadronic form 
factors does not exist.

From Table~\ref{tab:resonance_pdg} it is apparent that 
the values of photon helicity amplitudes $A_{1/2}$ 
and $A_{3/2}$ are unfortunately not accurate, in spite 
of the fact that the values directly control the 
magnitude of resonance contributions to the scattering
amplitude as can be clearly seen in Eq.~(\ref{eq:m_multipole}).
In order that the results of the present work do 
not to dramatically differ from those of our previous
analysis, where almost all resonance parameters were fixed to
the PDG values, in the first model (Model 1) we restrict 
the maximum variation of the photon amplitudes 
during the fitting process to 10\% 
of the original PDG values. In contrast to this, 
the variation of other parameters such as masses and 
total widths given in the Particle Data Book is mostly
smaller than 10\%. Therefore, in the latter we vary
the parameters within the allowed values given in the
Particle Data Book.

In the second model (Model 2) we do not constraint the variation
of the resonance parameters as strict as in Model 1, i.e.,
all parameters are allowed to vary within the PDG error
bars. Although we prefer Model 1 which retains the consistency
with our previous analysis, Model 2 will be useful in 
the present work once we want to investigate the model 
dependence of the mass determination of the narrow resonance
in the next section.

\begin{table}[!h]
\renewcommand{\arraystretch}{0.7}
  \centering
  \caption{The extracted coupling constants of the present work
    (Model 1 and Model 2) compared with those of 
    Kaon-Maid \cite{kaon-maid}. No hadronic form factors are
    used in both Model 1 and Model 2. The number of data points
    used in both models is 704.}
  \label{tab:numerical-result}
  \begin{ruledtabular}
  \begin{tabular}[c]{lrrr}
    Fit parameters & Model 1 & Model 2 & Kaon-Maid \\
\hline
  $g_{K \Lambda N}/\sqrt{4\pi}$&$-3.00$ &$-3.36$&$-3.80 $ \\
  $g_{K \Sigma N}/\sqrt{4\pi}$ &$ 1.30$ &$ 1.30$&$~~1.20$ \\
  $G^{V}_{K^{*}}/4\pi$         &$-0.46$ &$-0.45$&$-0.79 $ \\
  $G^{T}_{K^{*}}/4\pi$         &$ 0.48$ &$ 0.52$&$-2.63 $ \\
  $G^{V}_{K_1}/4\pi$           &$ 0.25$ &$ 0.32$&$~~3.81$ \\
  $G^{T}_{K_1}/4\pi$           &$-1.61$ &$-1.36$&$-2.41 $ \\
  $G_{Y^*}/\sqrt{4\pi}$        &$-1.71$ &$-1.06$&   -     \\
\hline
$\bvec{S_{11}(1650)}$\\
$M$ (MeV)        &1645 &1670 &- \\
$\Gamma$(MeV) &145 &164 &- \\
$A_{1/2}$($10^{-3}$ GeV$^{-1/2}$) &58 &69 &- \\
$\beta_K$  &0.031&0.031&   -     \\
$\phi$ (deg)  &176&195&   -     \\
\hline
$\bvec{D_{15}(1675)}$\\
$M$ (MeV)        &1680&1670&- \\
$\Gamma$(MeV)    &165 &134 &- \\
$A_{1/2}$($10^{-3}$ GeV$^{-1/2}$) &17 &13 &- \\
$A_{3/2}$($10^{-3}$ GeV$^{-1/2}$) &17 &24 &- \\
$\beta_K$  &0.019&0.010&   -     \\
$\phi$ (deg)  &11&34&   -     \\
\hline
$\bvec{F_{15}(1680)}$\\
$M$ (MeV)        &1680&1680&- \\
$\Gamma$(MeV)    &140 &140 &- \\
$A_{1/2}$($10^{-3}$ GeV$^{-1/2}$) &$-17$ &$-21$ &- \\
$A_{3/2}$($10^{-3}$ GeV$^{-1/2}$) &120 &121 &- \\
$\beta_K$  &0.000&0.000&   -     \\
$\phi$ (deg)  &185&219&   -     \\
\hline
$\bvec{D_{13}(1700)}$\\
$M$ (MeV)        &1750&1675&- \\
$\Gamma$(MeV)    &50 &82 &- \\
$A_{1/2}$($10^{-3}$ GeV$^{-1/2}$) &$-19$ &$-8$ &- \\
$A_{3/2}$($10^{-3}$ GeV$^{-1/2}$) &$-2$ &22 &- \\
$\beta_K$  &0.050&0.010&   -     \\
$\phi$ (deg)  &107&191&   -     \\
  \end{tabular}
  \end{ruledtabular}
\end{table}

\begin{table}[!h]
\renewcommand{\arraystretch}{0.7}
  \addtocounter{table}{-1}
  \centering
  \caption{The extracted coupling constants of the present work
    (Model 1 and Model 2) compared with those of 
    Kaon-Maid \cite{kaon-maid} (continued).}
  \begin{ruledtabular}
  \begin{tabular}[c]{lrrr}
    Fit parameters & Model 1 & Model 2 & Kaon-Maid \\
\hline
$\bvec{P_{11}(1710)}$\\
$M$ (MeV)        &1690&1699&- \\
$\Gamma$(MeV) &98&174&- \\
$A_{1/2}$($10^{-3}$ GeV$^{-1/2}$) &10&31&- \\
$\beta_K$  &0.140&0.140&   -     \\
$\phi$ (deg)  &88&98&   -     \\
\hline
$\bvec{P_{13}(1720)}$\\
$M$ (MeV)        &1700&1700&- \\
$\Gamma$(MeV)    &150&150&- \\
$A_{1/2}$($10^{-3}$ GeV$^{-1/2}$) &$16$ &18 &- \\
$A_{3/2}$($10^{-3}$ GeV$^{-1/2}$) &$-21$ &$-39$ &- \\
$\beta_K$  &0.048&0.048&   -     \\
$\phi$ (deg)  &195&191&   -     \\
\hline
  $\chi^2$   & 859 & 704 &-\\
  \end{tabular}
  \end{ruledtabular}
\end{table}

\subsection{Numerical results and comparison with data}

The numerical results of the fits are shown in 
Table~\ref{tab:numerical-result}, where the
background coupling constants of Kaon-Maid \cite{kaon-maid}
are also listed for comparison. Obviously,
the variation of the coupling constants
between Model 1 and Model 2 is less dramatic 
than between the two models and Kaon-Maid. 
Nevertheless, except for the $G_{K^*}^T$ coupling,
the sign of all coupling constants within the
three models is clearly consistent. In the
literature, the variation of these coupling
constants is a long standing problem. Although
Kaon-Maid was fitted to different experimental
data set and has different resonance configuration
as compared to the present work, Table~\ref{tab:numerical-result}
indicates that there is a tendency that the
variation starts to converge.

\begin{figure}[!t]
  \begin{center}
    \leavevmode
    \epsfig{figure=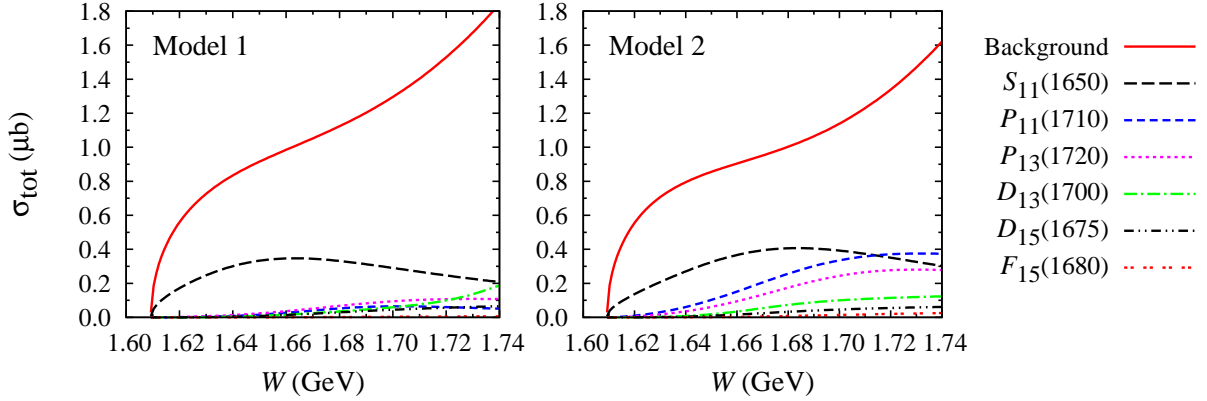,width=160mm}
    \caption{(Color online) Contribution of the background and  
      resonance terms to the total cross section for the two
      models used in the present work.}
   \label{fig:contrib} 
  \end{center}
\end{figure}

Contribution of the background and resonance terms 
in the two models are exhibited in Fig.~\ref{fig:contrib}.
It is obvious that the characteristic of resonance 
contributions can be comprehended from the values
of kaon branching ratio $\beta_K$ and photon 
helicity amplitudes $A_{1/2}$ and $A_{3/2}$ given in 
Table~\ref{tab:numerical-result}. The two models
show the same dominance of the background terms
and the same large contribution of the $S_{11}(1650)$ 
resonance. The differences between them 
appear at relatively higher energies. The 
background contribution of Model 2 is somewhat
suppressed at this kinematics in order to 
compensate contributions of the $S_{11}(1650)$, 
$P_{11}(1710)$ and $P_{13}(1720)$ resonances
that tend to increase. From Fig.~\ref{fig:contrib} 
(as well as Table~\ref{tab:numerical-result}) 
it is also seen that the peak of the $S_{11}(1650)$ 
contribution is shifted to higher energy in Model 2. 
It is obvious that Model 1 is more consistent with our previous 
multipole analysis \cite{Mart:2006dk}, i.e., the $S_{11}(1650)$ 
resonance contributes significantly, in contrast to 
the $P_{11}(1710)$. 

\begin{figure}[!h]
  \begin{center}
    \leavevmode
    \epsfig{figure=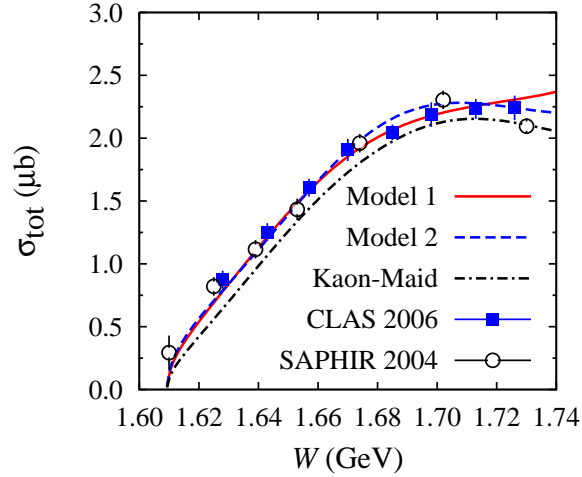,width=80mm}
    \caption{(Color online) Comparison between total cross sections 
      calculated from the Model 1, Model 2, 
      and Kaon-Maid \cite{kaon-maid} with the
      available experimental data from the SAPHIR \cite{Glander:2003jw}
      and  CLAS \cite{Bradford:2005pt} collaborations. 
      Note that error bars are statistical only and 
	all data shown in this figure
      were not used in the fitting process.
      }
   \label{fig:total} 
  \end{center}
\end{figure}

A comparison between the predicted total cross section 
from the two models as well as from the Kaon-Maid and
the available experimental data from SAPHIR 
\cite{Glander:2003jw} and CLAS \cite{Bradford:2005pt}
collaborations is shown in Fig.~\ref{fig:total}. It is
clear that both Model 1 and Model 2 show a better 
agreement than the Kaon-Maid, although at very high
energy ($\ge 1.730$ GeV) the total cross section predicted
by Model 1 starts to increase, in contrast to the
prediction of Model 2. This is understood from the
fact that at this energy regime contribution of 
the background terms 
in Model 1 is substantially larger than that in Model 2.

\begin{figure}[!ht]
  \begin{center}
    \leavevmode
    \epsfig{figure=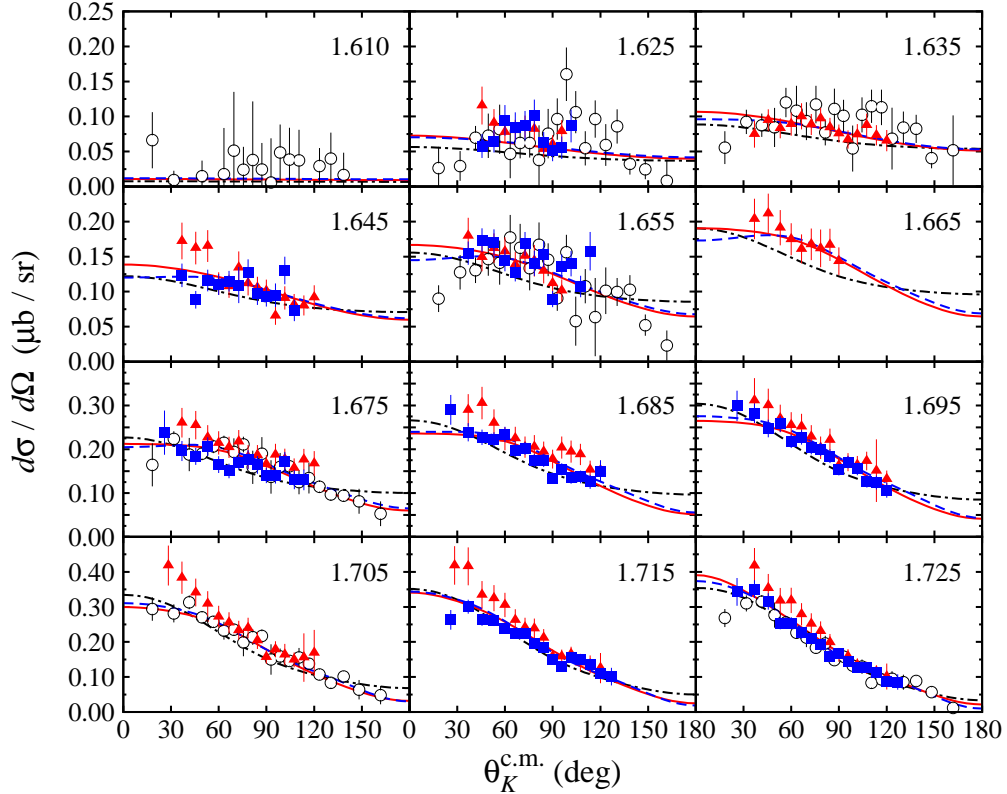,width=140mm}
    \caption{(Color online) Comparison between angular distributions of 
             differential cross section 
             calculated from the Model 1, Model 2, and Kaon-Maid 
             \cite{kaon-maid} with experimental data 
             from the SAPHIR (open circles)  \cite{Glander:2003jw}, 
             CLAS2006 (solid squares) \cite{Bradford:2005pt} 
	     and CLAS2010 (solid triangles) \cite{McCracken:2009ra} 
             collaborations.
             The corresponding total c.m. energy $W$ (in GeV) is
             shown in each panel. Experimental data displayed 
             in this figure were used
             in the fits. Notation of the curves is as in 
             Fig.~\ref{fig:total}.}
   \label{fig:dif_th} 
  \end{center}
\end{figure}

The angular distribution of the calculated differential
cross sections is exhibited in Fig.~\ref{fig:dif_th}.
Within the error bars of the available experimental data
we can say that all models work nicely in this case. 
Ideally, a full angular distribution of experimental 
data, such as the SAPHIR one, is desired for improving 
the model. Unfortunately, at forward angles 
SAPHIR data differ from CLAS data (see, e.g., panels with
$W=1.705$ and 1.715 GeV). Our models tend to approach the
SAPHIR data, presumably due to their smaller error bars.
In the case of Kaon-Maid model, the agreement with 
SAPHIR data is understandable because
the model was fitted to the previous version of SAPHIR 
data \cite{Tran:1998qw}, which are still consistent 
with the later version \cite{Glander:2003jw}.

\begin{figure}[!ht]
  \begin{center}
    \leavevmode
    \epsfig{figure=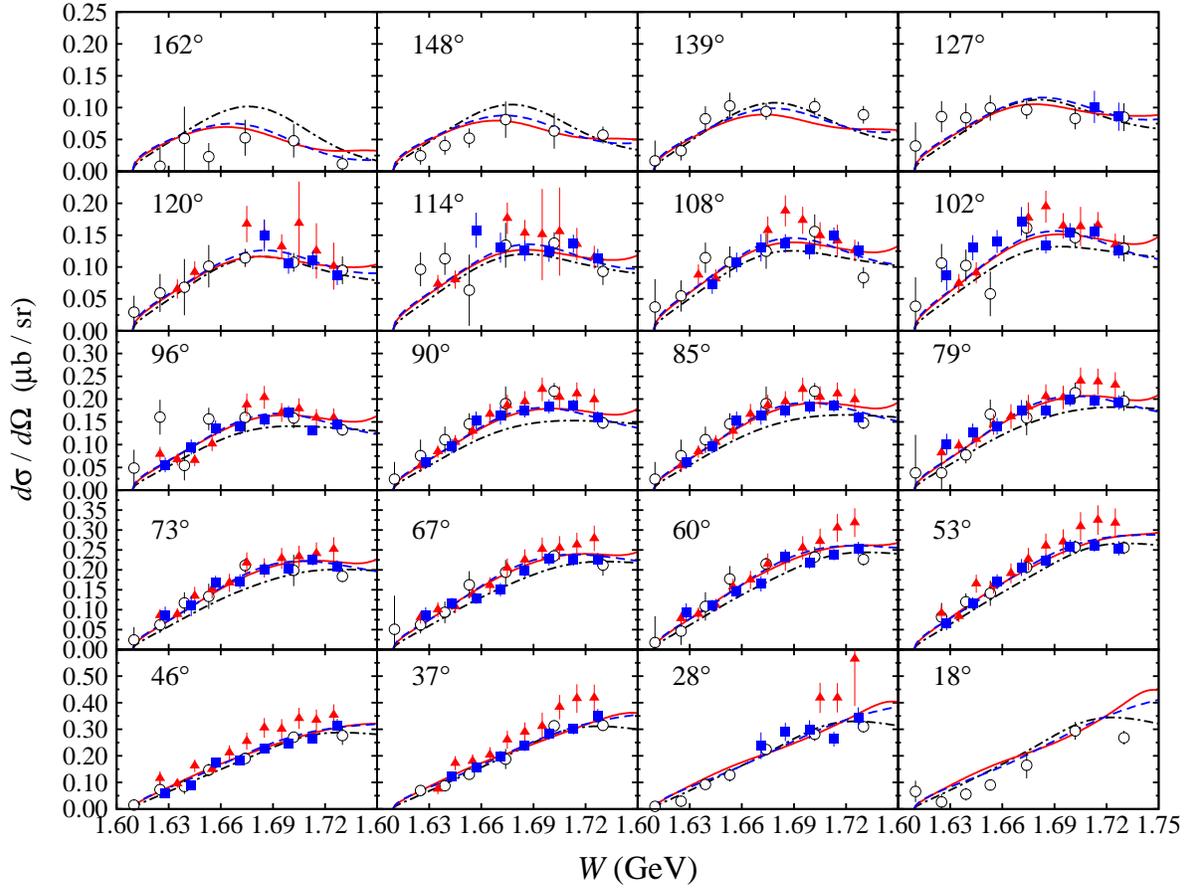,width=162mm}
    \caption{(Color online) As in Fig.~\ref{fig:dif_th}, 
             but for the total c.m. energy
             distribution. The corresponding kaon scattering 
             angle $\theta_K^{\rm c.m.}$ is shown in each panel. }
   \label{fig:dif_e} 
  \end{center}
\end{figure}

\begin{figure}[!ht]
  \begin{center}
    \leavevmode
    \epsfig{figure=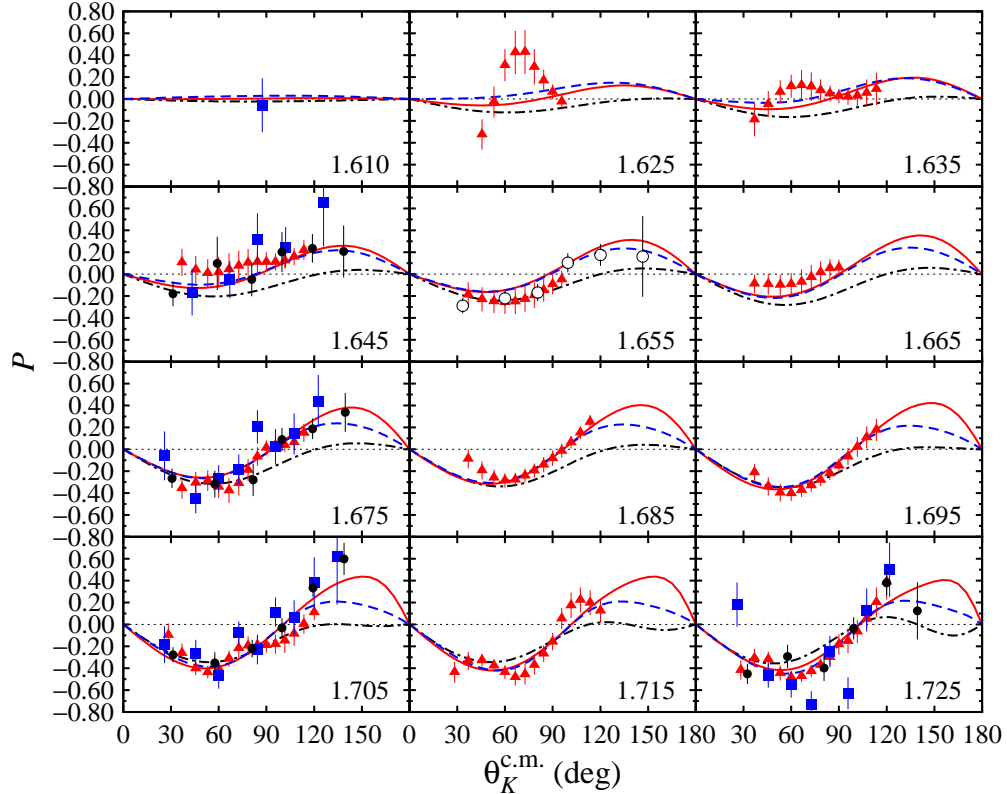,width=140mm}
    \caption{(Color online) Recoil polarization calculated from 
	     Model 1, Model 2, and Kaon-Maid \cite{kaon-maid} 
             compared with experimental data 
             from the SAPHIR \cite{Glander:2003jw} (open circles),
	     CLAS2006 (solid squares) \cite{Bradford:2005pt},
	     CLAS2010 (solid triangles) \cite{McCracken:2009ra},
             and GRAAL \cite{lleres07} (closed circles) collaborations.
             Notation of the curves is as in Fig. \ref{fig:total}.
             The corresponding total c.m. energy $W$ (in GeV) is
             shown in each panel.}
   \label{fig:polar} 
  \end{center}
\end{figure}

\begin{figure}[!ht]
  \begin{center}
    \leavevmode
    \epsfig{figure=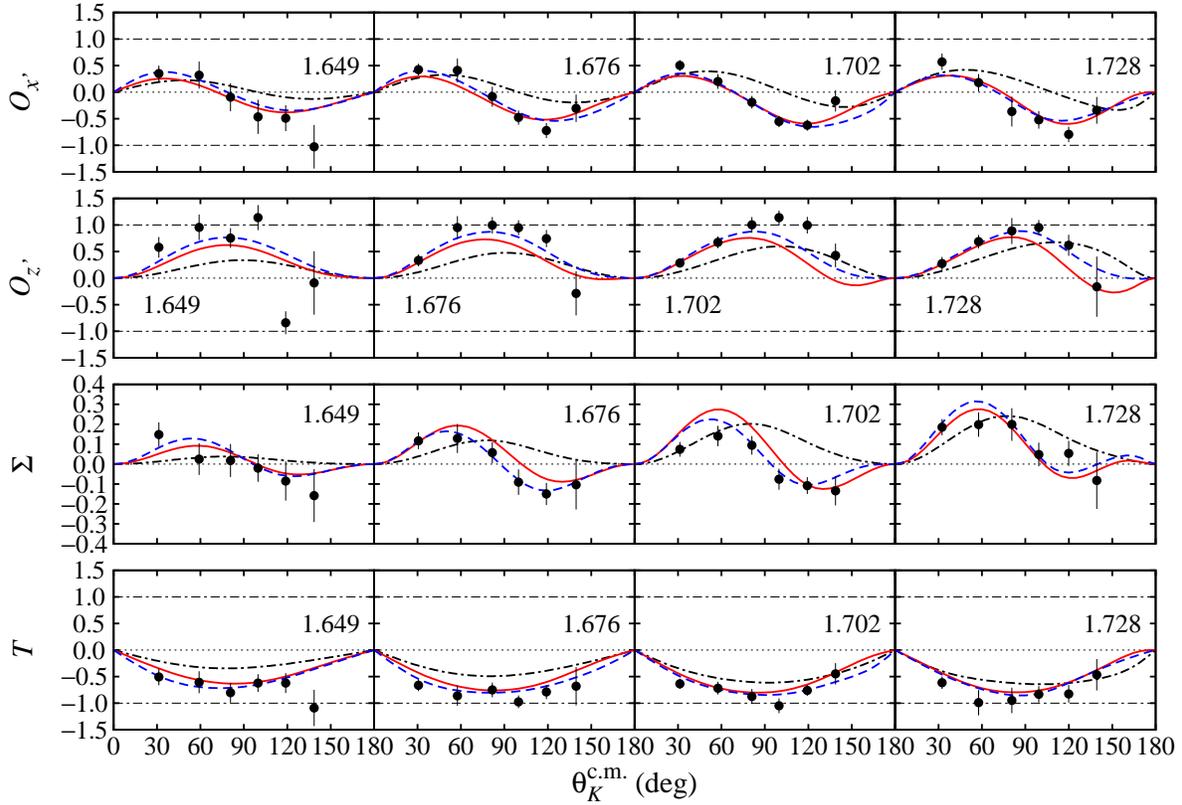,width=162mm}
    \caption{(Color online) The beam-recoil double polarization 
	observables $O_{x'}$ and $O_{z'}$, along with the target $T$ and
	photon $\Sigma$ asymmetries calculated from 
        Model 1, Model 2, and Kaon-Maid models \cite{kaon-maid} 
	compared with experimental data 
        from the GRAAL \cite{lleres09} collaboration.
             Notation of the curves is as in Fig. \ref{fig:total}.}
   \label{fig:ox_oz} 
  \end{center}
\end{figure}

\begin{figure}[!ht]
  \begin{center}
    \leavevmode
    \epsfig{figure=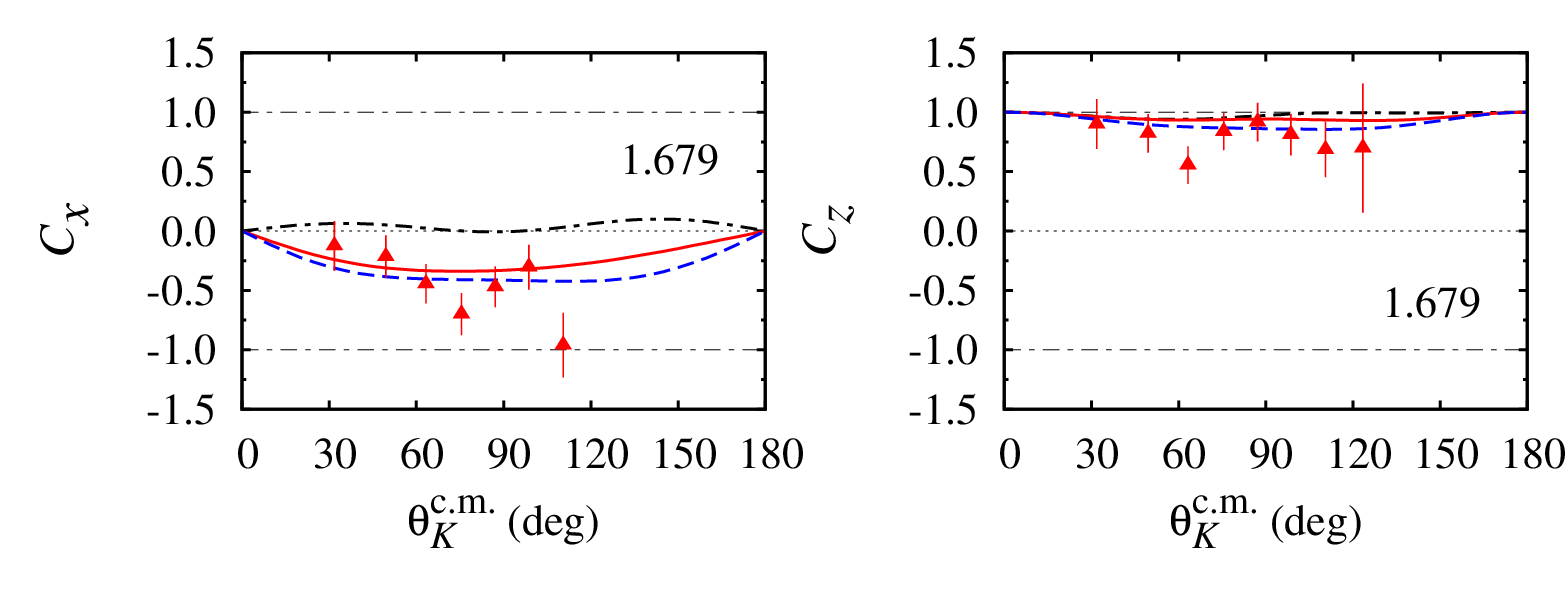,width=140mm}
    \caption{(Color online) The beam-recoil observables $C_x$ and $C_z$ 
	calculated from 
        Model 1, Model 2, and Kaon-Maid models \cite{kaon-maid} 
	compared with experimental data 
        from Ref.~\cite{Bradford:2006ba}. 
	Notation of the curves
      is as in Fig.~\ref{fig:total}.}
   \label{fig:cx_e} 
  \end{center}
\end{figure}

The energy distribution of differential cross sections for
20 different angle bins is shown in Fig.~\ref{fig:dif_e}. The
calculated cross sections of Model 1 and Model 2 are almost
identical except at very forward angles and near 
$\theta_K^{\rm c.m.}\approx 90^\circ$ with $W\ge 1.730$ GeV.
The agreement of both models with experimental data is
clearly better than in the case of Kaon-Maid. 
Although the experimental data do not show a clear 
resonance-like structure, a slight bump at 
$W\approx 1.690$ GeV can be observed. We note that within 
the error bars of the PDG resonance masses, all resonances 
considered in the present analysis could contribute to this 
bump. Besides that, the fact that threshold energies of 
all four $K\Sigma$ photoproductions are around 1.690 GeV, 
as shown in Table~\ref{tab:threshold}, could also be the
origin of this bump. 
Thus, we may conclude that an 
accurate extraction of resonance properties from kaon 
photoproduction at this energy point (1.690 GeV) would be 
a daunting task. The same situation could also happen at 
1.720 GeV, at which both $\rho p$ and $\omega p$ 
photoproduction have their thresholds.

\begin{table}[!ht]
  \centering
\caption{Threshold energies of meson photoproductions around
  1700 MeV in terms of the photon laboratory energy 
  ($E_\gamma^{\rm thr.}$) and the total c.m. energy ($W^{\rm thr.}$).
  Note that the threshold energy of kaon photoproduction 
  $\gamma +p\to K^+ + \Lambda$ is $E_\gamma^{\rm thr.}=911$ MeV
  ($W^{\rm thr.}=1609$ MeV).
  }
  \label{tab:threshold}
  \begin{ruledtabular}
  \begin{tabular}[c]{clccc}
    No.&Channel & $E_\gamma^{\rm thr.}$ (MeV)&~~~~~~&$W^{\rm thr.}$ (MeV)\\
    \hline
    1& $ \gamma + p$ $\longrightarrow$  $K^{+} + \Sigma^{0}$ &  1046 && 1686\\
    2& $ \gamma + p$ $\longrightarrow$  $K^{0} + \Sigma^{+}$ &  1048 && 1687\\
    3& $ \gamma + n$ $\longrightarrow$  $K^{+} + \Sigma^{-}$ & 1052  && 1691\\
    4& $ \gamma + n$ $\longrightarrow$  $K^{0} + \Sigma^{0}$ & 1051  && 1690\\
    5& $ \gamma + p$ $\longrightarrow$  $\rho + p$ & 1096  && 1714\\
    6& $ \gamma + p$ $\longrightarrow$  $\omega + p$ & 1109  && 1721\\
  \end{tabular}
  \end{ruledtabular}
\end{table}

The $\Lambda$ recoil polarization displayed in Fig.~\ref{fig:polar}
reveals an interesting phenomenon, especially at $W=1.625$ GeV
(see Fig.~\ref{fig:effect_polar} in the next section 
for the energy distribution of this structure).
The present analysis, as well as the Kaon-Maid model, cannot
reproduce the CLAS2010 data at this energy. We believe
that, assuming the data are accurate, such a structure cannot 
originate from an established nucleon resonance, since PDG
does not listed any single resonance at $W=1.625$ GeV. 
Since the polarization should be zero at threshold,
such an obvious structure 15 MeV above the threshold requires
special mechanism in the background terms that could
dramatically change the polarization slightly above the 
production threshold. The probability that a ''missing
resonance'' could solve this problem is very unlikely,
since the cross sections shown in Figs.~\ref{fig:total}
and \ref{fig:dif_e}
do not indicate any structure at this energy region.
At this stage we would just mention that 
future experimental and theoretical studies should 
address this problem as an important topic, since 
the polarization is automatically given in the 
kaon photoproduction experiments
and, on the other hand, problems at the production threshold 
can be better solved by a more consistent mechanism such as
chiral perturbation theory.

The photon-, target-, and double-polarization observables 
$O_{x'}$, $O_{z'}$,  $C_{x}$, $C_{z}$, 
are shown in Figs.~\ref{fig:ox_oz} and \ref{fig:cx_e}.
It is clear from these figures that both Model 1
and Model 2 can nicely describe the experimental data,
although we should admit that Model 2 can reproduce
the photon asymmetry data better than Model 1 due to its
smaller $\chi^2$.

\section{Search for a narrow resonance in kaon photoproduction}
\label{sec:search}

\begin{figure}[!ht]
  \begin{center}
    \leavevmode
    \epsfig{figure=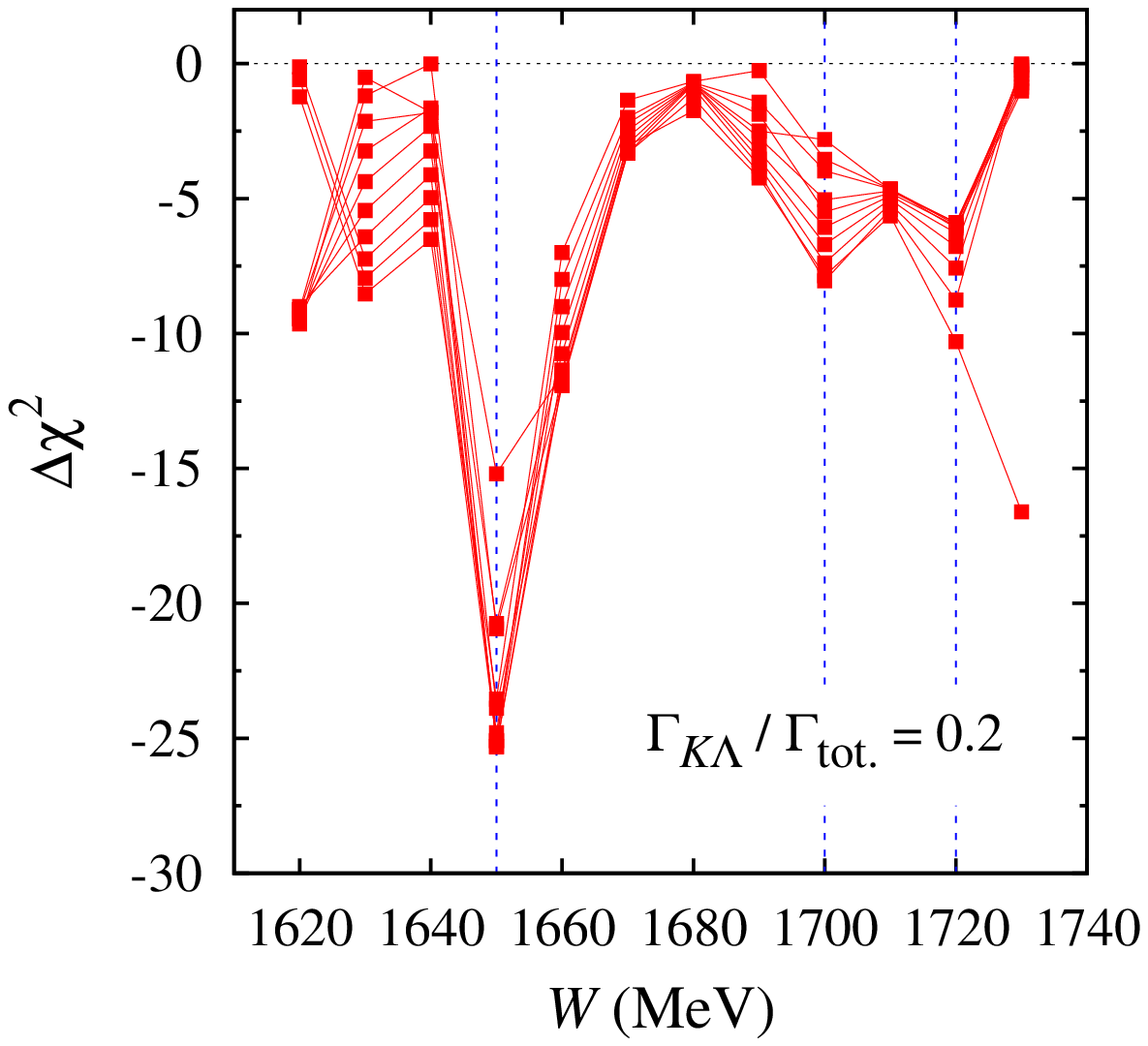,width=80mm}
    \caption{(Color online) Change of the $\chi^2$ in the
	fit of Model 1 due to the inclusion of the
	$P_{11}$ resonance with the mass scanned from 1620 
	to 1730 MeV (step 10 MeV) and $\Gamma_{\rm tot.}$ taken from
	1 to 10 MeV (step 1 MeV) for the $K\Lambda$ branching
	ratio 0.2. The three vertical lines indicate $m_{N^*}=1650$,
	1700 and 1720 MeV. A similar result is obtained
        for the $K\Lambda$ branching
	ratio 0.1 and 0.4. }
   \label{fig:scan-large} 
  \end{center}
\end{figure}

Having extended our photoproduction model up to $W=1730$ MeV,
we are ready now to study the possibility of observing a
narrow resonance in the $\gamma +p\to K^++\Lambda$ process.
As discussed in the Introduction, in Ref.~\cite{igor} an
attempt to find the existence of a narrow $J^p=1/2^+$
state was performed by including such a state in the
$\pi N$ partial wave $P_{11}$. The change of overall 
$\chi^2$ was scanned from $m_{N^*}=1610$ to 1760 MeV. In the
present work we follow this method, i.e., we include an extra narrow
$P_{11}$ resonance state in the kaon photoproduction amplitude
and scan the changes in the total $\chi^2$ after the inclusion,
in the energy range where our model is valid, i.e.,
$W=1620$ to 1730 MeV. Note that we do not start the scan
from $m_{N^*}=1610$ MeV, since the cross section at this threshold
energy is very small, while experimental data have very large
error bars (see the first panel of Fig.~\ref{fig:dif_th}).
As a consequence, predictions of our model at this kinematics would be less
reliable. The same behavior is also observed at the upper energy limit 
of the present analysis (1730 MeV).

\begin{figure}[!h]
  \begin{center}
    \leavevmode
    \epsfig{figure=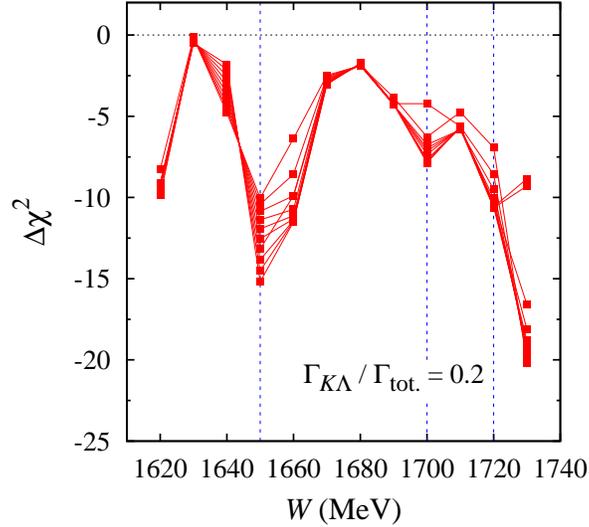,width=80mm}
    \caption{(Color online) As in Fig.~\ref{fig:scan-large}, but
	calculated by using total widths $\Gamma_{\rm tot.}$ from
	0.1 to 1 MeV (step 0.1 MeV).}
   \label{fig:scan-small} 
  \end{center}
\end{figure}

Figure \ref{fig:scan-large} displays the result of $\chi^2$ changes
($\Delta\chi^2$) after the inclusion of an extra $P_{11}$ resonance 
in Model 1 with the mass scanned
from 1620 to 1730 MeV, where the total width is varied from 1 to 10 MeV
with 1 MeV step. 
Although not shown in the figure, we have also
investigated the effect of the variation of $K\Lambda$ branching 
ratio and found only a small effect on the $\Delta\chi^2$. 
From Fig.~\ref{fig:scan-large} we can see that three minima appear
at $m_{N^*}=1650$, 1700, and 1720 MeV. 
Nevertheless, the minimum $\Delta\chi^2$ at
$m_{N^*}=1650$ MeV seems to be the most convincing one. 
For all values of the $K\Lambda$ branching ratio investigated
in this study the lowest values of
$\Delta\chi^2$ can be obtained by using $\Gamma_{\rm tot.}=5$ MeV.
Variation of the total width $\Gamma_{\rm tot.}$ results in a
variation of the $\Delta\chi^2$ value. 

As in Ref.~\cite{igor}, we have also repeated our calculation by using 
the total width varying from 0.1 to 1 MeV with 0.1 MeV step.
The result is shown in Fig.~\ref{fig:scan-small}. Although
the absolute values of $\Delta\chi^2$ obtained in this case 
are significantly 
different from those in the previous case, the similar pattern
still appears. Only at $m_{N^*}=1720$ MeV the minimum value of 
$\Delta\chi^2$ seems to disappear, since $\Delta\chi^2$ further
decreases at $m_{N^*}=1730$ MeV. This is due to the fact that 
our model is less reliable at the upper energy limit (see
total and differential cross sections shown in Figs.~\ref{fig:total}
and \ref{fig:dif_e}). As in the previous case, variation of 
the total width yields variation of the $\Delta\chi^2$ and
variation of the branching ratio changes this result slightly.
Note that the absolute values of $\Delta\chi^2$
here are smaller than in the case of larger $\Gamma_{\rm tot.}$. 
Therefore, at this stage we may conclude that our result prefers the 
total width values in the range of 
$1~{\rm MeV}\le\Gamma_{\rm tot.}\le 10~{\rm MeV}$.

\begin{figure}[!h]
  \begin{center}
    \leavevmode
    \epsfig{figure=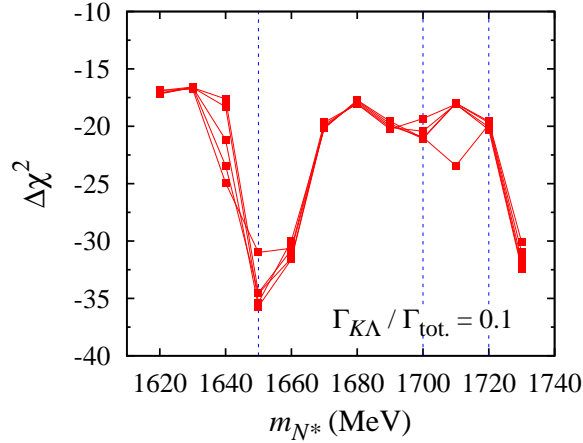,width=80mm}
    \caption{(Color online) As in Fig.~\ref{fig:scan-large}, but
	for Model 2 with the $K\Lambda$ branching ratio 0.1.}
   \label{fig:scan-model2} 
  \end{center}
\end{figure}

To investigate model dependence of our result in 
Fig.~\ref{fig:scan-model2} we display the same result
as in Fig.~\ref{fig:scan-large} with a branching ratio
of 0.1, but using Model 2. Once again, we see a similar
pattern found in Fig.~\ref{fig:scan-small}. We, therefore,
conclude that the minimum at $m_{N^*}=1650$ seems to be 
model independent, whereas the minima at $1700$ and 1720 MeV 
seem to disappear in Model 2. This phenomenon can be understood
from the fact that Model 2 has smaller $\chi^2$, i.e., the
agreement with experimental data is better than in Model 1.
Thus, improvement of the $\chi^2$ by adding nucleon
resonances is less likely in Model 2. As a consequence, the
number of minima in Model 2 is significantly reduced. To check
this argument, we have also analyzed a model that makes use of
nucleon resonances found in the analysis of new pion photoproduction
data from the CLAS collaboration \cite{dugger:2009}.
In this analysis only the $S_{11}(1650)$, $D_{15}(1675)$,
$F_{15}(1680)$, and $P_{13}(1720)$ states are considered,
whereas the variation of resonance parameters is very limited.
Obviously, the agreement with kaon photoproduction data
is worse ($\chi^2/N=1904$) than in Model 1 ($\chi^2=859$) 
or Model 2 ($\chi^2=704$). As a consequence, four minima 
are observed in the plot of  $\Delta\chi^2$, i.e., at
1650, 1670, 1690, and 1720 MeV, which clearly supports our argument
above.

\begin{figure}[!h]
  \begin{center}
    \leavevmode
    \epsfig{figure=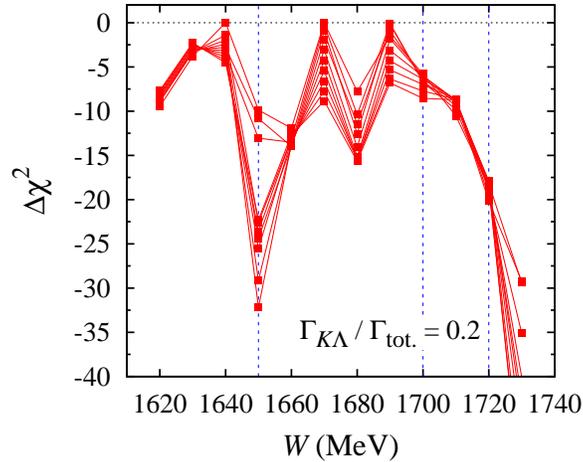,width=80mm}
    \caption{(Color online) Change of the $\chi^2$ in the
	fit of Model 1 due to the inclusion of the
	$S_{11}$ resonance with the mass scanned from 1620 
	to 1730 MeV (step 10 MeV) and using $\Gamma_{\rm tot.}$ from
	1 to 10 MeV (step 1 MeV) for the $K\Lambda$ branching
	ratio 0.2. As in Fig.~\ref{fig:scan-large}, 
	the three vertical lines indicate $m_{N^*}=1650$,
	1700 and 1720 MeV. Note that a new minimum
        at 1680 MeV appears in this case. }
   \label{fig:scan-S11} 
  \end{center}
\end{figure}

\begin{figure}[t]
  \begin{center}
    \leavevmode
    \epsfig{figure=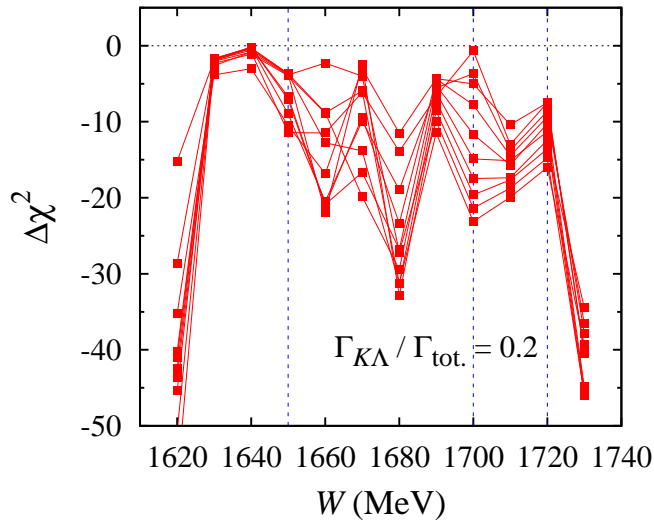,width=90mm}
    \caption{(Color online) As in Fig.~\ref{fig:scan-S11}, 
      but for the case of $P_{13}$. }
   \label{fig:scan-P13} 
  \end{center}
\end{figure}

Our finding corroborates the result of 
the topological soliton model of
Walliser and Kopeliovich~\cite{Walliser:2003dy}
discussed in the Introduction of this paper.
The $\chi$QSM of Diakonov and Petrov  \cite{diakonov2004} 
predicts a $J^p=1/2^+$ $N^*$ state
with a mass around 1650 MeV if the possible mixing 
between the lower-lying nucleonlike octet with the
antidecuplet is neglected. Thus, our result seems
to support this possibility. Although our finding
does not exclude the possibility that a narrow $P_{11}$ 
resonance with a mass of 1700 or 1720 MeV could exist, 
we believe that investigation of the resonance effects 
at these energies by using the present mechanism is 
difficult due to the opening of 
$K\Sigma$, $\rho p$, and $\omega p$  channels.

\section{Origin of the minima and possibility of other  resonance
states}
\label{sec:S11}

\begin{table}[b]
  \centering
\caption{Extracted narrow resonance parameters in the case
  that the resonance is an $S_{11}$, a $P_{11}$, or a 
  $P_{13}$ state. The $S_{11}$ and $P_{11}$ parameters
  are used in the following discussion, whereas the $P_{13}$
  parameters are given just for comparison.}
  \label{tab:narrow_res}
  \begin{ruledtabular}
  \begin{tabular}[c]{lccc}
    Extracted parameters & $S_{11}$ & $P_{11}$ & $P_{13}$ \\
    \hline
    $A_{1/2}~ (10^{-3}~{\rm GeV}^{1/2})$ &+90 &+40&$-$19 \\
    $A_{3/2}~ (10^{-3}~{\rm GeV}^{1/2})$ &- &-&+80 \\
    $m_{N^*}$ (MeV)& 1650 & 1650 & 1680 \\
    $\Gamma_{\rm tot.}$ (MeV) & 6 &5 & 8\\
    $\beta_K$ & 0.2 & 0.2& 0.2\\
    $\phi$ (deg.) &64 &67& 0 
  \end{tabular}
  \end{ruledtabular}
\end{table}

Although in this paper we have assumed the existence of a
narrow $P_{11}$ resonance as predicted by the $\chi$QSM 
and we only intent to explore the possibility
that it exists in the kaon photoproduction reaction, the results
found in the previous section could be also obtained  
by using other resonances, e.g., an $S_{11}$ or a $P_{13}$.
As discussed in the Introduction, a similar 
situation has been also found 
in the $\eta$ photoproduction~\cite{kim}.
To clarify this problem, in Fig.~\ref{fig:scan-S11} we
plot the changes of the $\chi^2$ if we replace the 
$P_{11}$ narrow resonance in the amplitude of
Model 1 with an $S_{11}$ ($J^p=1/2^-$) resonance. 
Obviously the same minimum at $m_{N^*}=1650$ MeV
is retained, but a new one clearly appears at 
$m_{N^*}=1680$ MeV. The appearance of the minimum $\Delta\chi^2$
at $m_{N^*}=1650$ MeV in all $\Delta\chi^2$ shown by 
Figs.~\ref{fig:scan-large} -- \ref{fig:scan-S11}
indicates that a real structure really exists at this energy,
although it is hardly seen in experimental data. 
However, the fact that both $S_{11}$ and $P_{11}$ could generate
this minimum means that a $J^p=1/2^-$
narrow resonance is also possible in the kaon photoproduction process.

Figure \ref{fig:scan-P13} displays the change of the $\chi^2$ 
in the fit of Model 1 if we include a $P_{13}$ ($J^p=3/2^+$)
narrow resonance 
instead of an $S_{11}$ or a $P_{11}$ state. Surprisingly, the
minimum at 1650 MeV almost vanishes and a clear minimum at
1680 MeV, as in the case of the $S_{11}$, appears. Besides that
we also observe two weaker minima at 1660 and 1700 MeV. However,
the minimum at 1680 MeV is interesting in this case, since 
the possibility that the structure found in the $\eta$ photoproduction
off a neutron can be explained by a $P_{13}$ resonance has been
discussed in Ref.~\cite{kuznetsov}. In fact, the most convincing
result with the smallest $\chi^2$ would be obtained if one 
used a $P_{13}(1685)$ state instead of a $P_{11}$ state \cite{kuznetsov}. 
Unfortunately, as discussed above and shown in Table~\ref{tab:threshold},
at energies around 1685 MeV there exists a number of meson 
photoproduction thresholds. Therefore, unless we could suppress 
the threshold effects at this energy point, further discussion of 
the $P_{13}(1685)$ would be meaningless at this stage.

\begin{figure}[t]
  \begin{center}
    \leavevmode
    \epsfig{figure=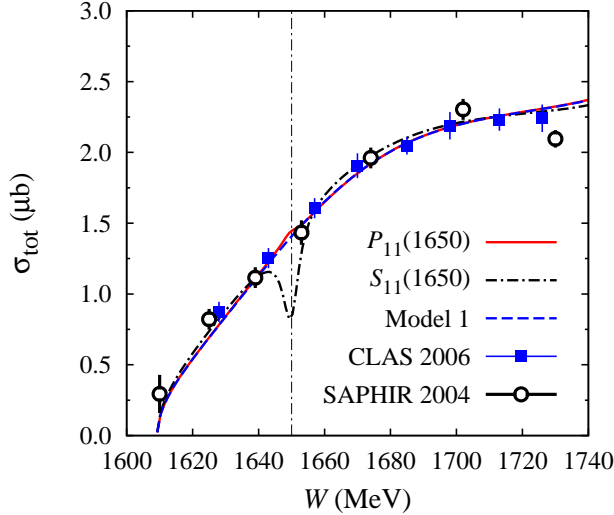,width=85mm}
    \caption{(Color online) Effects of the inclusion of 
	$P_{11}$ and $S_{11}$ resonances  with resonance 
        parameters given in Table~\ref{tab:narrow_res}
        on the total cross 
	section of the $\gamma+p\to{K^+}+\Lambda$ process. }
   \label{fig:effect1} 
  \end{center}
\end{figure}

In the PWA it is possible to check
whether the true resonance extracted from the analysis is
a $P_{11}$ or not. A true resonance would yield an effect
only when inserted into the correct partial amplitude \cite{igor}.
In the present analysis such a technology is unfortunately not
available. However, in principle, different natures of the $P_{11}$, 
$S_{11}$, and $P_{13}$ resonances represented by their different formulations
are traceable in the measured observables.

\begin{figure}[t]
  \begin{center}
    \leavevmode
    \epsfig{figure=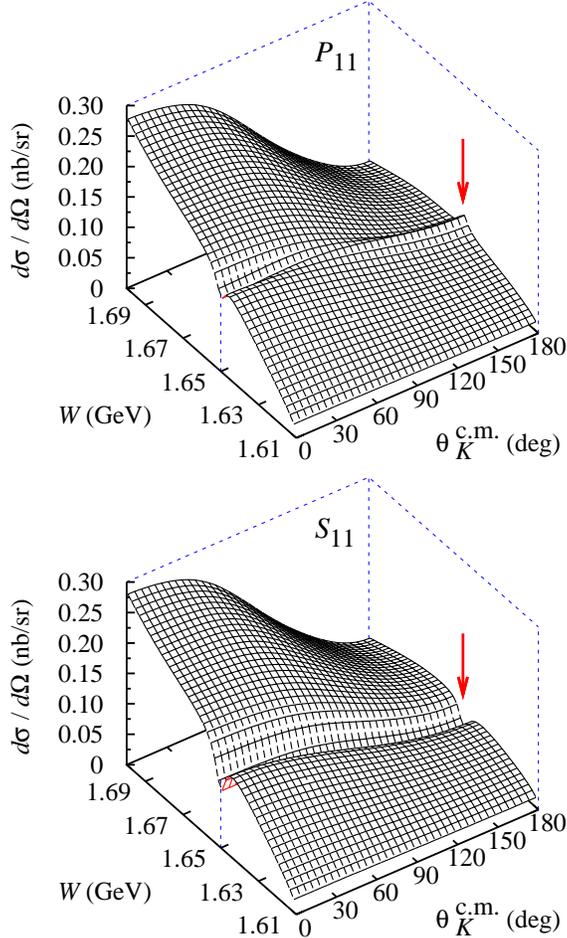,width=80mm}
    \caption{(Color online) Effects of the inclusion of the $P_{11}$ 
      (top figure) and $S_{11}$ (bottom figure) resonances with 
      resonance parameters given in Table~\ref{tab:narrow_res} on the 
      differential cross sections. Arrows in the figures
      indicate the position of $W=1.650$ GeV.}
   \label{fig:effect2} 
  \end{center}
\end{figure}

To prove this argument, in Fig.~\ref{fig:effect1} 
we show different effects
generated by the $P_{11}$ and the $S_{11}$ resonances
on the total cross section compared with experimental
data. Note that we use the resonance parameters
given in Table~\ref{tab:narrow_res}, which are
obtained as the best fits to experimental data.
Interestingly, the $S_{11}$ resonance generates
a clear dip at $W=1650$ MeV  
in the total cross section, whereas the effect
of the $P_{11}$ resonance is almost negligible.
Clearly, such a dip is not observed by the
presently available data, since the energy bin
of the data is larger than the width of the dip. 
Should the structure around 1650 MeV really exist, then 
future experiments with smaller energy bins 
(e.g. 2 MeV) would be required to resolve it 
and, simultaneously, to single out the true resonance.

How can the structure predicted by the $\Delta\chi^2$ minima 
in Figs.~\ref{fig:scan-large} -- \ref{fig:scan-S11}
almost disappear in the total cross section? The answer is
given in Fig.~\ref{fig:effect2}, where we can see 
that the effect of the $P_{11}$ resonance on the 
differential cross sections is in fact comparable 
with that of the $S_{11}$ resonance, but the effect 
changes almost drastically at $\theta_K^{\rm c.m.}
\approx 70^\circ$ from decreasing to increasing
cross section as the kaon angle increases. This phenomenon obviously 
disappears in the total cross section after an angular 
integration over all possible angles averages this effect.
In the case of the $S_{11}$ we obtain a decreasing effect
in the whole angular distribution, which therefore produces 
an obvious dip at $W=1650$ MeV in the total cross section.

\begin{figure}[t]
  \begin{center}
    \leavevmode
    \epsfig{figure=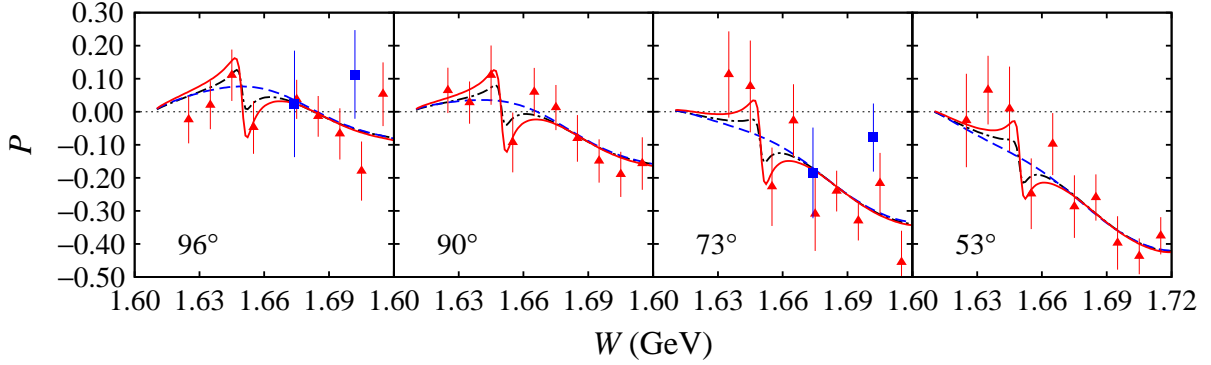,width=160mm}
    \caption{(Color online) As in Fig.~\ref{fig:effect1}, but for the
      $\Lambda$ recoil polarization. Notation for the experimental
    data is as in Fig.~\ref{fig:polar}.}
   \label{fig:effect_polar} 
  \end{center}
\end{figure}

It is obviously important to know which data are really responsible
for the minimum at 1650 MeV as shown in Figs.~\ref{fig:scan-large} -- 
\ref{fig:scan-S11}. For this purpose, we have scrutinized contributions 
of individual data to the $\chi^2$ in our fits
and found that this minimum originates mostly from the $\Lambda$ recoil 
polarization data as displayed in Fig.~\ref{fig:effect_polar}.
From this figure we can see that there exists a dip at $W\approx 1650$ MeV
in the whole angular distribution of data. It is also
apparent that both $P_{11}$ and $S_{11}$ states can nicely reproduce
the dip. Therefore, it seems to us that the recoil polarization is 
not the suitable observable to distinguish the possible states at
1650 MeV. Nevertheless, more precise recoil polarization data 
are still urgently required in order to support the finding
of the present work as well as to remove uncertainties in 
the position of the dip. Definitely, JLab FROST project looks
promising for this purpose \cite{frost}.

In the beam-recoil double polarization observables $O_{x'}$ and $O_{z'}$
the presence of an $S_{11}$ or a $P_{11}$ narrow state predicts a
small structure at 1650 MeV. In the case of the $S_{11}$ state 
the structure is slightly more
obvious than in the case of the $P_{11}$ state. We note that for the
$O_{x'}$ observable this structure increases as the kaon angle increases.
The opposite behavior is shown by the $O_{z'}$ observable. 
Although the structures seem to be mild, their differences 
generated by the different natures of the $S_{11}$ and $P_{11}$ 
resonances might provide important 
observables to determine the origin of the $\Delta\chi^2$ minimum at
1650 MeV.

\begin{figure}[t]
  \begin{center}
    \leavevmode
    \epsfig{figure=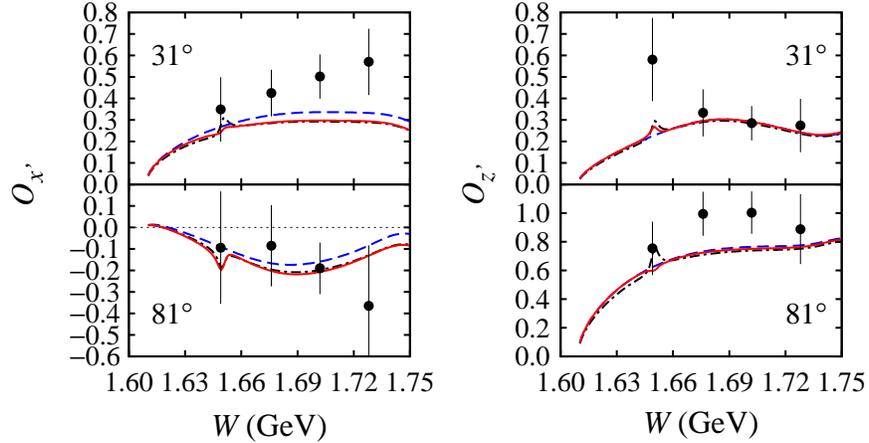,width=120mm}
    \caption{(Color online) As in Fig.~\ref{fig:effect1}, but for the
      $O_{x'}$ and $O_{z'}$ double polarization observables.
      Experimental data are from the GRAAL \cite{lleres09} collaboration.}
   \label{fig:effect_ox_oz} 
  \end{center}
\end{figure}

\begin{figure}[t]
  \begin{center}
    \leavevmode
    \epsfig{figure=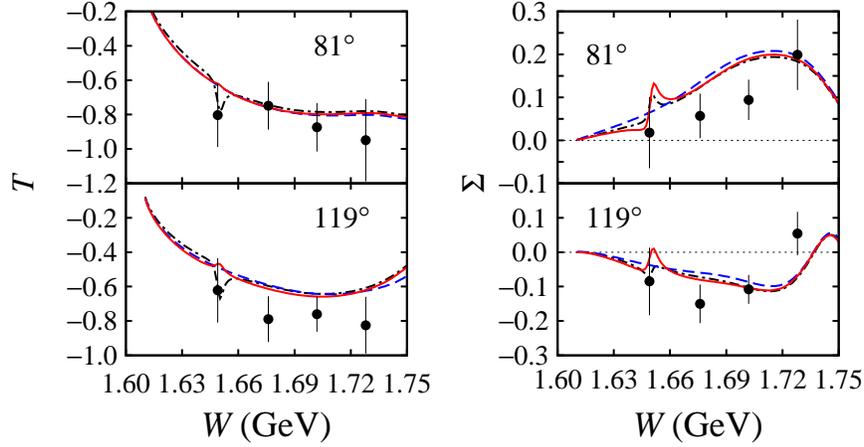,width=120mm}
    \caption{(Color online) As in Fig.~\ref{fig:effect_ox_oz}, but for the
      target and photon beam asymmetry.}
   \label{fig:effect_target} 
  \end{center}
\end{figure}

Although the effect is milder, the same behavior is also 
shown by the target asymmetry 
$T$, as exhibited in Fig.~\ref{fig:effect_target}. In the case of photon
asymmetry, given in the same figure, the effect generated by 
the $P_{11}$ states is more obvious than that by the $S_{11}$ state. 

\begin{figure}[!h]
  \begin{center}
    \leavevmode
    \epsfig{figure=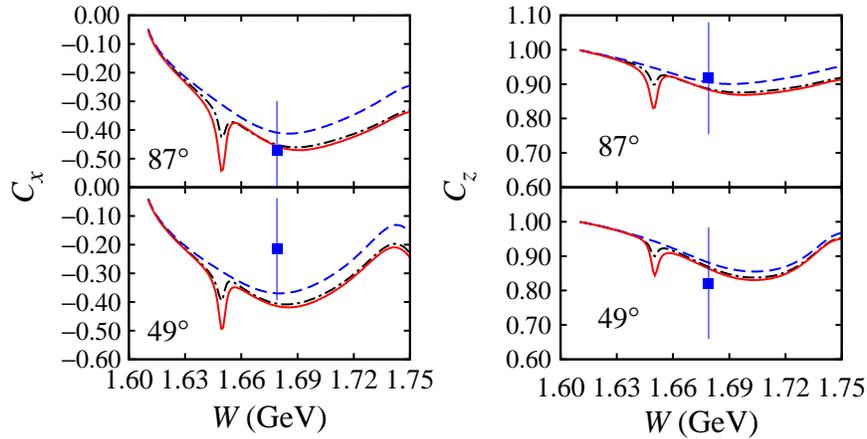,width=120mm}
    \caption{(Color online) As in Fig.~\ref{fig:effect1}, but for the
      $C_{x}$ and $C_{z}$ double polarization observables.
    Experimental data are from Ref.~\cite{Bradford:2006ba}.}
   \label{fig:effect_cx_cz} 
  \end{center}
\end{figure}

The effect of the narrow resonances is also found to be sizable
on the beam-recoil double polarization observables $C_x$ and $C_z$,
as displayed in Fig.~\ref{fig:effect_cx_cz}. Different from the photon
or beam-recoil asymmetries, $\Sigma$ and $O_{z'}$, 
here we observe that both resonances 
yield a clear dip at $W=1650$ MeV. Although  $C_x$ and $C_z$ 
probably cannot
distinguish the effects of $S_{11}$ and $P_{11}$ states, the sizable 
dips produced here indicate that these 
observables are seem to be promising for investigation of the
narrow resonance existence in kaon photoproduction.

We believe that new measurements with the present 
accelerator and detector technologies would be able to resolve the effects 
shown in Figs.~\ref{fig:effect1} and \ref{fig:effect2}.
Furthermore, more precise kaon photoproduction
data with energy bins about 2 MeV would be already able to
discriminate the effect of $P_{11}$ and $S_{11}$
resonances on the total cross section and 
improve the accuracy of our calculation.

\section{Summary and conclusion}
\label{sec:summary}
We have studied the possibility of observing the $J^p=1/2^+$
narrow resonance, the nonstrange member of antidecuplet
baryons predicted by the $\chi$QSM, in kaon
photoproduction off a proton. For this purpose, we have
constructed two isobar models that can reproduce
experimental data from threshold up to $W=1730$ MeV,
based on our previous effective Lagrangian model.
After inserting the resonance in the models we
analyzed the changes in the total $\chi^2$ 
with the variation of $m_{N^*}$ from 1620 to
1730 MeV and 
found the most convincing minimum at $m_{N^*}=1650$ MeV.
This finding is observed for all isobar models used
in this investigation 
and could be distinguished from the $J^p=1/2^-$ and
$3/2^+$ resonances, provided that more precise kaon photoproduction
data were available. Furthermore, our conclusion does not
change with the variation of the total width and $K\Lambda$
branching ratio of the resonance. Although the mass 
of the resonance obtained in our calculation (i.e., 1650 MeV) 
is slightly different from those obtained from 
the $\pi N$ and $\eta N$ reactions, the 1650 MeV mass 
corroborates the result of the topological soliton model
and the  
calculation utilizing the Gell-Mann-Okubo 
rule without mixing between the lower-lying nucleonlike 
octet with the antidecuplet. Needless to say
that more precise kaon photoproduction data
are crucial to prove and improve our 
present calculation.

\section*{Acknowledgment}
The author thanks Igor I. Strakovsky for 
suggesting this study and for fruitful
discussions.
Supports from the University of Indonesia
and the Competence Grant of the Indonesian 
Ministry of National Education are gratefully
acknowledged.

\end{document}